\documentclass[smallcondensed]{svjour3}
\usepackage{amsmath}
\usepackage{amsfonts}
\usepackage{amssymb}
\usepackage{graphicx}

\usepackage{color}

\def\b{\begin{equation}}
\def\e{\end{equation}}
\def\ba{\begin{eqnarray}}
\def\ea{\end{eqnarray}}

\def\gabf{\mbox{\boldmath{$\gamma$}}}

\def\u{\uparrow}
\def\d{\downarrow}
\def\le{\langle}
\def\re{\rangle}

\begin{document}

\title{The role of spin in entanglement generated by expanding spacetime}

\author{Roberto Pierini \and Shahpoor Moradi  \and Stefano Mancini}

\institute{R. Pierini \at
             School of Science and Technology, University of Camerino, 62032 Camerino, Italy \\
             INFN Sezione di Perugia, 06123 Perugia, Italy
              \email{roberto.pierini@unicam.it}          
           \and
           S. Moradi \at
   University of Calgary, Department of Geoscience, Calgary, Canada  \\
        \email{moradis@ucalgary.ca}
       \and
           S. Mancini \at
School of Science and Technology, University of Camerino, 62032 Camerino, Italy \\
             INFN Sezione di Perugia, 06123 Perugia, Italy
\email{stefano.mancini@unicam.it}
}

\date{Received: date / Accepted: date}


\maketitle

\begin{abstract}
We investigate the effects of spin on entanglement arising in Dirac field
in an expanding spacetime characterized by the Robertson-Walker metric.
We present a general approach that allows us to treat the case where only
 charge conservation is required,
 as well as the case where also angular momentum conservation is required.
 We find that in both situations entanglement originated from the vacuum and quantified by subsystem entropy
behaves the same and does not qualitatively deviates from the spinless case. Differences only arise when particles and/or antiparticles are present in the input state.
\end{abstract}


\section{Introduction}

In recent years we witnessed a continuous tendency of quantum information theory to permeate relativistic
physics (see \cite{PT04} for a review on seminal works along this line and \cite{Ivet-rev,exp} for more recent achievements).
In particular, much attention has been devoted to entanglement characterization \cite{Horo} in relativistic frameworks
starting from special relativity \cite{JSS07,CCM12} and moving on to general relativity \cite{FS05,AL06,M09,L09}.
When dealing with curved spacetime it is customary to require flatness in remote past and far future in order to unambiguously define particles and antiparticles (often the conformally flat Robertson-Walker metric is employed for such purpose) \cite{bdv}.
Taking this approach it has been possible to learn about certain aspects
of entanglement in curved spacetime \cite{TU04,Shi04,ball,st,iv}.
In particular, the long known phenomenon of particle-antiparticle creation from vacuum has been studied to  the end of characterizing entanglement in Refs.\cite{iv,MMM12}.
In such a context, although Dirac fields were analyzed beside bosonic fields, the role of spin has not yet clarified.
A first analysis including spin was carried out in Ref. \cite{MPM14} by assuming angular momentum conservation besides charge conservation.

Here we start from a more general scenario where only charge conservation is required
and we explicitly find the unitary representation of the involved Bogolyubov transformations.
With this approach we are also able to survey the scenario where both angular
momentum and charge conservation are required (treated differently in \cite{MPM14}).
It results that in both cases entanglement originated from vacuum and quantified by subsystem entropy
behaves the same and does not qualitatively deviates from the spinless case.
This holds true also when considering two spin particles and two spin antiparticles in input.
Moreover, entanglement is not created for input charge equal to 2, while it is created for input charge equal to 1 but only in the presence of spin.
Finally, a remarkable difference between the cases of charge conservation and charge and momentum conservation only appears for having one particle and one anti-particle in the input state.

The paper is organized as follows. In Section \ref{expand}
we recall the Dirac spinors in expanding spacetime and present the most general
Bogolyubov transformations connecting remote past and far future (input and output respectively)
in Section \ref{Bogo}.
In Section \ref{Ccons} we construct from the given Bogolyubov transformations an input vacuum
resulting to only respect charge conservation and evaluate the entanglement generated starting from it.
Then, in Section \ref{CLcons} we argue the Bogolyubov transformations
leading to an input vacuum that shows angular momentum conservation besides charge conservation.
At the end also the spinless case is revisited (Section \ref{entaC}). 
Although, for the sake of simplicity, the main body of the paper focuses on the input vacuum, in Section \ref{conclu} we draw our conclusion by comparing the results obtainable for various (pure and factorable) input states.
In Appendix \ref{unirep} the unitary representation of Bogolyubov transformations is discussed.
The corresponding unitary operator is worked out in Appendix \ref{Uact}.
Finally Appendix \ref{Excited} contains the expressions of subsystem entropies when starting from excited input states.


\section{Dirac spinors in expanding spacetime}\label{expand}

We start by considering the Robertson-Walker line element (see e.g. \cite{bdv})
\begin{equation*}
ds^2=a^2(\tau)(-d\tau^2+dx^2+dy^2+dz^2),
\label{ds}
\end{equation*}
where the dimensionless
conformal time $\tau$ is related to cosmological time $t$ by
$\tau=\int{a^{-1}(t)dt}$. Here $a(\tau)$ is the scale factor determining the spacetime expansion rate. We assume flat spacetime as $\tau\rightarrow \pm\infty$.
The covariant generalization of Dirac equation for spin-$\frac{1}{2}$ field $\psi$ of mass $m$ is given by
\b
\left(\tilde{\gamma}^{\mu}(\partial_{\mu}+\Gamma_{\mu})
+m\right)\psi=0,
\label{Deq}
\e
where the curved gamma matrices
$\tilde{\gamma}^{\mu}$ are related to flat ones through
$\tilde{\gamma}^{\mu}:=a^{-1}\gamma^{\mu}$ and
\begin{equation*}
\Gamma_{\mu}=\frac{1}{4}\frac{\dot{a}}{a}[\gamma_{\mu},\gamma_0],
\label{Gm}
\end{equation*}
are the spin connections.
Here and below dot  denotes the derivative with respect to conformal time $\tau$.

Writing $\psi =a^{-3/2}(\gamma^{\nu}\partial_{\nu}-M)\varphi$, with $M=ma(\tau)$,
in \eqref{Deq} we get
\b
g^{\mu\nu}\partial_{\mu}\partial_{\nu}\varphi-
\gamma^{0}\dot{M}\varphi-M^2\varphi=0,
\label{newDeq}
\e
being $g^{\mu\nu}$ the flat metric as opposed to the actual spacetime metric
$\tilde{g}^{\mu\nu}:=a^{-2}g^{\mu\nu}$.
Next, given the flat spinors $u_d$ and $v_d$ (with $d=\u,\d$) satisfying the relations
\begin{equation*}
\gamma^0u_d=-iu_d, \quad\quad \gamma^0v_d=iv_d,
\label{gammarel}
\end{equation*}
we set
\begin{equation}
\begin{split}
\varphi^{(-)}&:=N^{(-)}f^{(-)}(\tau)u_d
e^{ i{\mathbf{p}}\cdot{\mathbf{x}}}, \\
\varphi^{(+)}&:=N^{(+)}f^{(+)}(\tau) v_d e^{ i{\mathbf{p}}\cdot{\mathbf{x}}},
\label{varphidef}
\end{split}
\end{equation}
with ${\mathbf{x}}$, ${\mathbf{p}}$ position and momentum vectors in $\mathbb{R}^3$.
The functions $f^{(\pm)}$ obey the differential equation
\begin{equation}
\ddot{f}^{(\pm)}+
\left(|{\mathbf{p}}|^{2}+M^2\pm
i\dot{M}\right)f^{(\pm)}=0.
\label{feq}
\end{equation}
Let us define by $f^{(\pm)}_{in/out}$ the solutions behaving
as positive frequency modes with respect to conformal
time~$\tau$ near the asymptotic past (\emph{in}) / future (\emph{out}),
i.e. $\dot{f}^{(\pm)}_{in/out}(\tau)\approx - iE_{in/out}f^{(\pm)}_{in/out}(\tau)$ with
\begin{equation*}
E_{in/out}\equiv\sqrt{|{\mathbf{p}}|^{2}+M_{in/out}^{2}},\quad M_{in/out}\equiv m a(\tau\to-/+\infty).
\label{Einout}
\end{equation*}

Then we can introduce spinors that behave like positive and negative energy spinors, respectively, in the asymptotic regions:
\begin{align}
\begin{split}
U({\mathbf{p}},d,{\mathbf{x}},\tau)&:=N(\gamma^{\nu}\partial_{\nu}-M)f^{(-)}(\tau)u_de^{i{\mathbf{p}}\cdot{\mathbf{x}}},
\nonumber\\
V({\mathbf{p}},d,{\mathbf{x}},\eta)&:=N(\gamma^{\nu}\partial_{\nu}-M) f^{(+)^*}(\tau)v_de^{-i{\mathbf{p}}\cdot{\mathbf{x}}},\nonumber
\label{UVpd}
\end{split}
\end{align}
with the normalization constant $N=\frac{1}{\sqrt{2M(E+M)}}$ and $^*$ denoting the complex conjugation.


\section{Input-Output transformations}\label{Bogo}

Let us assume $\{U_{in},V_{in}\}$ and $\{U_{out},V_{out}\}$ to be two
complete sets of mode solutions for the Dirac equation \eqref{newDeq} which define
particles and anti-particles in asymptotic regions and have corresponding
vacua, $|0\re_{in}$ and $|0\re_{out}$ respectively.
 Physically, $|0\re_{in}$
is the state with no incoming particles (anti-particles) in remote
past and $|0\re_{out}$ is the state with no outgoing
particles (anti-particles) in the far future. When $\tau\rightarrow
\pm\infty$, the spacetime is flat and
the dynamics of the field is that of the free field. So we have two natural
quantizations of the field, associated with two Fock spaces.
The Dirac field operator can hence be written as
\begin{equation}
\begin{split}
\psi({\mathbf{x}},\eta)&= \int d{\mathbf{p}}\sum_{d}\left[a_{in}({\mathbf{p}},d)
U_{in}({\mathbf{p}},d,\eta)e^{i{\mathbf{p}}\cdot{\mathbf{x}}}+b^{\dag}_{in}({\mathbf{p}},d)V_{in}({\mathbf{p}},d,\eta)e^{-i{\mathbf{p}}\cdot{\mathbf{x}}}\right]\\
&=\int d{\mathbf{p}}\sum_{d}\left[a_{out}({\mathbf{p}},d)
U_{out}({\mathbf{p}},d,\eta)e^{i{\mathbf{p}}\cdot{\mathbf{x}}}+b^{\dag}_{out}({\mathbf{p}},d)V_{out}({\mathbf{p}},d,\eta)
e^{-i{\mathbf{p}}\cdot{\mathbf{x}}}\right],
\label{psi-inout}
\end{split}
\end{equation}
 where $a_{in},b_{in}$ and  $a_{out},b_{out}$
 are annihilation operators of particles and anti-particles in the $in$ and $out$ asymptotic regions respectively.
Actually $a_{in}$  and $b_{in}$ differ from
$a_{out}$  and $b_{out}$ because
 they do not correspond to physical particles outside the
 $in$ region. However, it is possible to relate the operators of
$in$-particles to those of $out$-particles by Bogolyubov transformation
\cite{bdv}.

We can find the Bogolyubov transformations relating the input regions with the output in the following way. Properties of the equation (\ref{feq}) give the following relation for its solutions 
\begin{equation}
\begin{split}
f^{(-)}_{in}(\tau)&=A^{(-)}(p)f^{(-)}_{out}(\tau)+B^{(-)}(p){f^{(+)}}^*_{out}(\tau),
\\
f^{(+)}_{in}(\tau)&=A^{(+)}(p)f^{(+)}_{out}(\tau)+B^{(+)}(p){f^{(-)}}^*_{out}(\tau).
\end{split}
\label{finp}
\end{equation}
where $p\equiv|\mathbf{p}|$, and $A^{\pm}$ and $B^{\pm}$ are the Bogolyubov coefficients.

For $\tau\rightarrow\infty$ Eq.\eqref{finp} reads
\b
\begin{split}
f^{(-)}_{in}\rightarrow A^{(-)}(p )e^{-iE_{out}}+B^{(-)}(p )e^{iE_{out}}\\
f^{(+)}_{in}\rightarrow A^{(+)}(p )e^{-iE_{out}}+B^{(+)}(p )e^{iE_{out}}
\label{afinp}
\end{split}
\e
Inserting \eqref{afinp} in \eqref{psi-inout} and equating the coefficients of curved-space spinors we arrive at
\b
\begin{split}
a_{in}({\bf p},d)&={\cal A}^*({p}) a_{out}({\bf p},d)-\sum_{d'}\beta^*_{dd'}({-\bf p}) b_{out}^\dag (-{\bf p},d'),\\
b_{in}^\dag({-\bf p},d)&=\sum_{d'}\beta_{d'd}({-\bf p}) a_{out}({\bf p},d')+ {\cal A}( p) b_{out}^\dag (-{\bf p},d)  \,,
\end{split}
\label{Bab}
\e
where
\begin{equation}
\begin{split}
{\cal A}(p)&=\sqrt{\frac{E_{out}}{E_{in}}\frac{E_{out}+M_{out}}{E_{in}+M_{in}}}A^{(-)}(p )\,,
\end{split}
\end{equation}
and
\begin{equation}
\begin{split}
\beta_{dd'}({\mathbf{p}})&=-i\frac{v^{\dag}_{d'}\, \gabf \cdot{\mathbf{p}}\,u_d\,B^{(-)}(p)}
{\sqrt{\frac{E_{in}}{E_{out}}(E_{out}+M_{out})(E_{in}+M_{in})}},
\end{split}
\label{bero}
\end{equation}
Note that $\beta_{dd'}(-{\bf p})=-\beta_{dd'}({\bf p})$ . The corresponding inverse transformations are
\b
\begin{split}
a_{out}({\bf p},d)&={\cal A}(p) a_{in}({\bf p},d)+\sum_{d'}\beta^*_{dd'}({-\bf p}) b_{in}^\dag (-{\bf p},d'),\\
b_{out}^\dag(-{\bf p},d)&=-\sum_{d'}\beta_{d'd}(-{\bf p}) a_{in}({\bf p},d')+ {\cal A}^*( p) b_{in}^\dag (-{\bf p},d)\,.
\end{split}
\label{Babinv}
\e
Hereafter, without loss of generality, the quantity $\mathcal{A}( p)$ will be considered real.
The expectation value of the $out$ number of particles in the $in$ vacuum (i.e., number of created particles) with momentum
${\mathbf p}$ and spin projection $d$ is
\b
n^p({\mathbf p},d):=\le 0_{in}
|a^{\dag}_{out}({\mathbf{p}},d)a_{out}({\mathbf{p}},d)|0_{in}
\re =\sum_{d'} |\beta_{d,d'}(-{\mathbf{p}})|^2,
\label{np}
\e
where the adjoint of \eqref{Bab} has been used.
Similarly, the expectation value of $out$ number of anti-particles in the $in$ vacuum (i.e.,
number of created anti-particles) is
\b
n^a(-{\mathbf{p}},d):=\le 0_{in}|
b^{\dag}_{out}(-{\mathbf{p}},d)b_{out}(-{\mathbf{p}},d)|0_{in}
\re=\sum_{d'} |\beta_{d',d}(-{\mathbf{p}})|^2.
\label{na}
\e
Then, the total number of created particles with momentum ${\mathbf{p}}$ (total number density) results
by combining \eqref{np} and \eqref{na} and summing up over spin projections
\b
\label{densityn}
n({\mathbf{p}}):=\sum_d\ \left[ n^p({\mathbf{p}},d)+n^a(-{\mathbf{p}},d)\right].
\e
From now on, to simplify the notation, we will omit the argument of Bogolyubov coefficients by assuming it to always be $\mathbf{p}$,  unless otherwise stated. We are legitimate in doing that because of the absence of mode mixing in the Bogolyubov transformations \eqref{Bab} and  \eqref{Babinv}. Also, we assume that ladder operators $a$ $(a^{\dagger})$ always refer to particles with momentum ${\bf p}$, while operators labeled by $b$ $(b^{\dagger})$ always refer to anti-particles with momentum ${-\bf p}$, unless otherwise stated.

\section{Charge conservation}\label{Ccons}

The unitary operator (acting on the Fock space) that represents the transformation \eqref{Bab} has been derived in Appendix \ref{unirep}.
By applying it to the $out$ vacuum we will get the $in$ vacuum, that is
$|0_p;0_{-p}\re_{in}=\mathcal{U} |0_p;0_{-p}\re_{out}$. This notation refers to the fact that if an observer in the input region detects some vacuum state, an observer in the output region will detects some ``evolved" state which depends on the dynamics of the universe.
It results in (see Appendix \ref{Uact})
\begin{align}
|0_p;0_{-p}\re_{in}=&{\mathcal{A}}^2\Big(|0_p;0_{-p}\re_{out}-\frac{\beta^*_{\u\d}}{{\mathcal{A}}}|\u_p;\d_{-p}\re_{out}
-\frac{\beta^*_{\d\u}}{{\mathcal{A}}}|\d_p;\u_{-p}\re_{out}  \nonumber \\
-&\frac{\beta^*_{\u\u}}{{\mathcal{A}}}|\u_p;\u_{-p}\re_{out}-\frac{\beta^*_{\d\d}}{{\mathcal{A}}}|\d_p;\d_{-p}\re_{out}+
\frac{\beta^*_{\d\u}\beta^*_{\u\d}-\beta^*_{\u\u}\beta^*_{\d\d}}{{\mathcal{A}}^2}|\u\d_p,\u\d_{-p}\re_{out}\Big) \,.
\label{1inout}
\end{align}
This is the way a vacuum state at $in$ is seen at $out$, i.e. no longer as a vacuum.
Notice that the states at the r.h.s. have zero net charge likewise the one at l.h.s., however the angular momentum is non zero at the r.h.s. (differently from the l.h.s.) due to the presence of  states like $|\d_p;\d_{-p}\re_{out}$
and $|\u_p;\u_{-p}\re_{out}$.
The particle anti-particle density operator corresponding to \eqref{1inout} in the $out$ region reads
\begin{align}
\label{rhooutcre1}
\varrho^{(out)}_{p,-p}=\frac{1}{2}{\mathcal{A}}^4 \Big[&
|0_p;0_{-p}\re\le 0_p;0_{-p}|
-2\frac{\beta_{\u\d}}{\mathcal{A}}
 |0_p;0_{-p}\re\le\u_p;\d_{-p}| \nonumber\\
& -2\frac{\beta_{\d\u}}{\mathcal{A}}
 |0_p;0_{-p}\re\le\d_p;\u_{-p}|-2\frac{\beta_{\u\u}}{\mathcal{A}}
 |0_p;0_{-p}\re\le\u_p;\u_{-p}|\nonumber\\
&-2\frac{\beta_{\d\d}}{\mathcal{A}}
 |0_p;0_{-p}\re\le\d_p;\d_{-p}|
  +2 \frac{\beta_{\u\d} \beta_{\d\u}-\beta_{\u\u} \beta_{\d\d}}{{\mathcal{A}}^2}
 |0_p;0_{-p}\re\le \u\d_p;\u\d_{-p}|\nonumber\\
 &+\frac{|\beta_{\u\d}|^2}{\mathcal{A}^{2}}
 |\u_p;\d_{-p}\re\le\u_p;\d_{-p}|
+2\frac{\beta_{\u\d}^*\beta_{\d\u}}{{\mathcal{A}}^{2}}
 |\u_p;\d_{-p}\re\le\d_p;\u_{-p}|\nonumber\\
&+2\frac{\beta_{\u\d}^*\beta_{\u\u}}{{\mathcal{A}}^{2}}
|\u_p;\d_{-p}\re\le\u_p;\u_{-p}|   
+2\frac{\beta_{\u\d}^*\beta_{\d\d}}{{\mathcal{A}}^{2}}
 |\u_p;\d_{-p}\re\le\d_p;\d_{-p}|\nonumber\\
&-2 \beta_{\u\d}^{*}\frac{\beta_{\u\d} \beta_{\d\u}-\beta_{\u\u} \beta_{\d\d}}{\mathcal{A}^{3}}
 |\u_p;\d_{-p}\re\le \u\d_p;\u\d_{-p}|
+\frac{|\beta_{\d\u}|^2}{\mathcal{A}^{2}}
 |\d_p;\u_{-p}\re\le\d_p;\u_{-p}|\nonumber\\
&+2\frac{\beta_{\d\u}^*\beta_{\u\u}}{{\mathcal{A}}^{2}}
|\d_p;\u_{-p}\re\le\u_p;\u_{-p}|
+2\frac{\beta_{\d\u}^*\beta_{\d\d}}{{\mathcal{A}}^{2}}
 |\d_p;\u_{-p}\re\le\d_p;\d_{-p}|\nonumber\\
&-2 \beta_{\d\u}^{*}\frac{\beta_{\u\d} \beta_{\d\u}-\beta_{\u\u} \beta_{\d\d}}{\mathcal{A}^{3}}
 |\d_p;\u_{-p}\re\le \u\d_p;\u\d_{-p}|
+\frac{|\beta_{\u\u}|^2}{\mathcal{A}^{2}}
 |\u_p;\u_{-p}\re\le\u_p;\u_{-p}|\nonumber\\
&+2\frac{\beta_{\u\u}^*\beta_{\d\d}}{{\mathcal{A}}^{2}}
|\u_p;\u_{-p}\re\le\d_p;\d_{-p}| 
-2 \beta_{\u\u}^{*}\frac{\beta_{\u\d} \beta_{\d\u} 
-\beta_{\u\u}\beta_{\d\d}}{\mathcal{A}^{3}}
|\u_p;\u_{-p}\re\le \u\d_p;\u\d_{-p}| \nonumber\\
&+\frac{|\beta_{\d\d}|^2}{\mathcal{A}^{2}}
 |\d_p;\d_{-p}\re\le\d_p;\d_{-p}| 
-2 \beta_{\d\d}^{*}\frac{\beta_{\u\d} \beta_{\d\u}-\beta_{\u\u} \beta_{\d\d}}{\mathcal{A}^{3}}|\d_p;\d_{-p}\re\le \u\d_p;\u\d_{-p}| \nonumber\\
&+\frac{(|\beta_{\d\d}|^2+|\beta_{\u\d}|^2)^2}{\mathcal{A}^{2}}
 |\u\d_p;\u\d_{-p}\re\le\u\d_p;\u\d_{-p}|\Big]+ {\rm h.c.}.
\end{align}
The reduced density operator for particle is obtained by tracing over antiparticle $\varrho_p^{(out)}={\rm Tr}_{-p}\left(\varrho_{p,-p}^{(out)}\right)$ and it results
\begin{align}
\varrho_p^{(out)}=&
{\mathcal{A}}^4 |0_p\re\le 0_p|
+{\mathcal{A}}^{2}( |\beta_{\u\d}|^2+|\beta_{\u\u}|^2) |\u_p\re\le\u_p|
+{\mathcal{A}}^{2} (|\beta_{\d\u}|^2+|\beta_{\u\u}|^2) |\d_p\re\le\d_p|\nonumber\\
&+ (|\beta_{\u\d}|^2 +|\beta_{\u\u}|^2)^2 |\u\d_p\re\le\u\d_p|   
+{\mathcal{A}}^2(\beta_{\u\u}^*\beta_{\d\u}+\beta_{\u\d}^*\beta_{\d\d})|\u_p\re\le\d_p| \nonumber\\
&+{\mathcal{A}}^2(\beta_{\d\u}^*\beta_{\u\u}+\beta_{\u\d}\beta_{\d\d}^*)|\d_p\re\le\u_p| \,.
\label{reducedrho1}
\end{align}
The latter can be simplified by accounting for the fact that
\b\label{relBogfinal}
(\beta_{\u\u}^*\beta_{\d\u}+\beta_{\u\d}^*\beta_{\d\d})=(\beta_{\d\u}^*\beta_{\u\u}+\beta_{\u\d}\beta_{\d\d}^*)=0 \,,
\e
as shown in Appendix \ref{unirep}.
In the case of only charge conservation, the Bogolyubov coefficients are related to the total number of particles $n$ (see Eqs. \eqref{np}, \eqref{na} and \eqref{densityn}) in the following way
\b
\label{An1}
{\cal A}^2=\frac{4-n}{4},\quad
|\beta_{\u\d}|^2 +|\beta_{\u\u}|^2 = \frac{n}{4}.
\e

To evaluate the amount of particle-antiparticle entanglement of \eqref{rhooutcre1} we can use the subsystem entropy \cite{Horo} (because the state is pure)
\b
\label{entropydef}
S\left(\varrho_p^{(out)}\right)\equiv -{\rm Tr} \left( \varrho_p^{(out)}\log_2 \varrho_p^{(out)} \right).
\e
That leads to  (together with relations \eqref{relBogfinal} and \eqref{An1})
\begin{align}
\label{Sspincreation1}
S\left(\varrho^{(out)}_{p}\right)=-2\left(\frac{4-n}{4}\right) \log_2\left(\frac{4-n}{4}\right)
-2\left(\frac{n}{4}\right) \log_2\left(\frac{n}{4}\right) \,.
\end{align}
It is worth remarking that \eqref{Sspincreation1} is a concave function of $n$ taking minimum value (zero) for $n=0$ and $n=4$ and maximum value (two) for $n=2$.


\section{Charge and angular momentum conservation}\label{CLcons}

To have angular momentum conservation the coefficients of $|\d_p;\d_{-p}\re_{out}$ and $|\u_p;\u_{-p}\re_{out}$
in \eqref{1inout} would have be zero,  however, based on \eqref{bero}, $\beta_{\uparrow\uparrow}$ and $\beta_{\downarrow\downarrow}$ are non zero.
This means that we have to remove them from
\eqref{Bab} which simplifies to (see also \cite{MPM14})
\b
\begin{split}
a_{in}(d)&={\cal A}^* \, a_{out}(d)+\beta^*_{d,-d} \, b_{out}^\dag (-d),\\
b_{in}^\dag(d)&=-\beta_{-d,d}\, a_{out}(-d)+ {\cal A} \, b_{out}^\dag (d)  \,.
\end{split}
\label{Babsimple}
\e
The unitary operator acting on the Fock space and representing such transformation in derived in the Appendix \ref{unirep}.  Applying such unitary on the $out$ vacuum gives (see Appendix \ref{Uact})
\b
\label{0inout}
|0_p;0_{-p}\re_{in}={\mathcal{A}}^2\left(|0_p;0_{-p}\re_{out}-\frac{\beta^*_{\u\d}}{{\mathcal{A}}}|\u_p;\d_{-p}\re_{out}
-\frac{\beta^*_{\d\u}}{{\mathcal{A}}}|\d;\u_{-p}\re_{out}+
\frac{\beta^*_{\u\d}\beta^*_{\d\u}}{{\mathcal{A}}^2}|\u\d_p,\u\d_{-p}\re_{out}\right).
\e
Again this is the way a vacuum state at $in$ is seen at $out$, i.e. no longer as a vacuum.
Notice however that this time the net angular momentum of the state at the r.h.s. is zero likewise that of the state at the l.h.s. The particle anti-particle density operator corresponding to \eqref{0inout} in the $out$ region reads
\begin{align}
\label{rhooutcre2}
\varrho^{(out)}_{p,-p}=\frac{1}{2}\Big[&
{\mathcal{A}}^4 |0_p;0_{-p}\re\le 0_p;0_{-p}|
-2{\mathcal{A}}^{3}\beta_{\u\d}
 |0_p;0_{-p}\re\le\u_p;\d_{-p}| \nonumber\\
& -2{\mathcal{A}}^{3}\beta_{\d\u}
 |0_p;0_{-p}\re\le\d_p;\u_{-p}|
  +2{\mathcal{A}}^{2} \beta_{\u\d} \beta_{\d\u}
 |0_p;0_{-p}\re\le \u\d_p;\u\d_{-p}|\nonumber\\
 &+{\mathcal{A}}^{2} |\beta_{\u\d}|^2
 |\u_p;\d_{-p}\re\le\u_p;\d_{-p}|
 +2{\mathcal{A}}^{2} \beta_{\u\d}^*\beta_{\d\u}
 |\u_p;\d_{-p}\re\le\d_p;\u_{-p}|\nonumber\\
&  -2{\mathcal{A}} |\beta_{\u\d}|^2 \beta_{\d\u}
 |\u_p;\d_{-p}\re\le\u\d_p;\u\d_{-p}|
+{\mathcal{A}}^{2} |\beta_{\d\u}|^2
 |\d_p;\u_{-p}\re\le\d_p;\u_{-p}|\nonumber\\
 & -2{\mathcal{A}} \beta_{\u\d} |\beta_{\d\u}|^2
|\d_p;\u_{-p}\re\le\u\d_p;\u\d_{-p}|
+ |\beta_{\u\d}|^2 |\beta_{\d\u}|^2
|\u\d_p;\u\d_{-p}\re\le\u\d_p;\u\d_{-p}| \Big] + {\rm h.c.}.
\end{align}
The reduced particle state results
\b
\varrho_p^{(out)}=
{\mathcal{A}}^4 |0_p\re\le 0_p|
+{\mathcal{A}}^{2} |\beta_{\u\d}|^2 |\u_p\re\le\u_p|
+{\mathcal{A}}^{2} |\beta_{\d\u}|^2 |\d_p\re\le\d_p|
+ |\beta_{\u\d}|^2 |\beta_{\d\u}|^2 |\u\d_p\re\le\u\d_p|.
\label{reducedrho0}
\e
Notice that from \eqref{np} we have
\b
|\beta_{\u\d}|^2= n^p(\d), \quad |\beta_{\d\u}|^2= n^p(\u).
\e
Then, referring to \eqref{densityn} and assuming
$n^p(\u)=n^p(\d)=n^a(\u)=n^a(\d)=n/4$, we obtain the coefficients appearing in \eqref{reducedrho0}
solely depending on $n$, that is
\b
\label{An}
{\cal A}^2=\frac{4-n}{4},\quad
|\beta_{\u\d}|^2 = |\beta_{\d\u}|^2 = \frac{n}{4}.
\e
To evaluate the amount of particle-antiparticle entanglement of \eqref{rhooutcre2} we use again the subsystem entropy \eqref{entropydef} obtaining
\b
\label{Sspincreation}
S\left(\varrho^{(out)}_{p}\right)=
-2\left(\frac{4-n}{4}\right) \log_2\left(\frac{4-n}{4}\right)
-2\left(\frac{n}{4}\right) \log_2\left(\frac{n}{4}\right) \,,
\e
which is exactly the same expression of \eqref{Sspincreation1},
although the states \eqref{rhooutcre1} and \eqref{rhooutcre2} are not the same.


\section{Entanglement in $1+1$ spacetime}\label{entaC}

In the case of $1+1$ spacetime all coefficients of one particle and one antiparticle states
in \eqref{1inout} would be the same due to the lack of dependence on spin indexes.
Moreover the coefficients of two particle and two antiparticle states would be zero.
This means that \eqref{Bab} simplifies to
\b
\begin{split}
a_{in}&= {\cal A} \, a_{out}
+ \beta^{\ast} \, b^{\dag}_{out} \\
b^{\dag}_{in} &=
-\beta \, a_{out}+{\cal A}\,b^{\dag}_{out}.
\end{split}
\label{ainoutspinless}
\e
The corresponding unitary operator is derived in the Appendix \ref{unirep}.
Once it is applied to the $out$ vacuum it yields (see Appendix \ref{Uact})
\ba
|0_p;0_{-p}\re_{in}={\cal A}\left(|0_p;0_{-p}\re_{out}-\frac{\beta^*}{\cal A}|1_p;1_{-p}\re_{out}\right).
\label{invac}
\ea
The density operator in the $out$ region corresponding to \eqref{invac} reads
\b
\varrho^{(out)}_{p,-p}=\frac{1}{2}\Big[
{{\cal A}^2} |0_p;0_{-p}\re \le 0_p;0_{-p}|
-2\beta {\cal A}  |0_p;0_{-p}\re\le 1_p;1_{-p} |
+|\beta|^2  |1_p;1_{-p}\re \le 1_p;1_{-p}|\Big] +{\rm h.c.},
 \label{rhoinvac}
\e
and the reduced particle state is
\b
\varrho^{(out)}_{p}=
{{\cal A}^2} |0_p\re \le 0_p|
+|\beta|^2  |1_p\re \le 1_p|.
\e
Referring to again to \eqref{densityn} we can now assume
$n^p=n^a=n/2$ and
Eq.\eqref{An} becomes
\b
\label{An2}
{\cal A}^2=\frac{2-n}{2},\quad
|\beta|^2 = \frac{n}{2}.
\e
Then, the amount of entanglement  of \eqref{rhoinvac}, thanks to \eqref{entropydef}
and \eqref{An2}, results
\b
S\left(\varrho^{(out)}_{p}\right)=-\left(1-\frac{n}{2}\right)\log_2\left(1-\frac{n}{2}\right)-\frac{n}{2}\log_2\frac{n}{2} .
\label{Spa}
\e
Also \eqref{Spa} results a concave function of $n$ taking however minimum value (zero) for $n=0$ and $n=2$ and maximum value (one) for $n=1$.

Comparing \eqref{Sspincreation1} and \eqref{Sspincreation} with \eqref{Spa} we end up with the following relation
\begin{equation*}
S_{spin}(n)=2S_{spinless}(n/2),\quad 0 \leq n\leq 4.
\end{equation*}


\section{Conclusion}\label{conclu}

In conclusion, we have investigated the role of spin on particle-antiparticle entanglement arising from the vacuum of the Dirac field
in an expanding spacetime characterized by the Robertson-Walker metric.
We have considered two approaches in defining the input vacuum in terms of output states:
one requiring only charge conservation,
the other where both charge and angular momentum conservation was required.
We have found that, although the resulting output states are different in the two approaches,
their amount of entanglement is the same and simply related to that of spinless case, namely
$S_{spin}(n)=2S_{spinless}(n/2)$, for $0 \leq n\leq 4$ the density of particle creation.
We can then conclude that, in this framework, spin does not play a relevant role in entanglement generation. 
More generally this is not true. When considering input states different from the vacuum it is possible to find qualitative and quantitative differences for the sub-system entropy. Its expression for various input states is reported in Appendix \ref{Excited}. 
Upon making a comparison, we can summarize here the following results:
\begin{itemize}
\item[i)]
For charge equal to 2, i.e. two spin particles or two spin antiparticles, it is $S_{spin}(n)=0$.
\item[ii)]
For charge equal to 1 it is $S_{spin}(n)=-\left(\frac{4-n}{4}\right)\log_2\left(\frac{4-n}{4}\right)-\left(\frac{n}{4}\right)\log_2\left(\frac{n}{4}\right)$, while $S_{spinless}(n/2)=0$.
\item[iii)]
For zero charge with two spin particles and two spin antiparticles, the situation is identical to 
that of vacuum, where $S_{spin}(n)=2S_{spinless}(n/2)$.
\item[iv)] For zero charge with one spin particle and one spin antiparticle we have:
\begin{itemize}
\item[a)] in case of only charge conservation $S_{spin}$ becomes function of an additional parameter $\lambda$, besides $n$, in a way depending on whether the spin are parallel or anti-parallel;
\item[b)] in case of charge and momentum conservation $S_{spin}$ only depends on $n$ and it is 
$S_{spin}(n)=2S_{spinless}(n)$ for anti-parallel spin, while $S_{spin}(n)=0$ for parallel spin.
\end{itemize} 
\end{itemize}
Hence, provided that the input state is not the vacuum, by measuring entanglement we could distinguish between different scenarios.    

Finally, spin may play some role on entanglement degradation once entanglement between modes of different momentum is considered (as pointed out in \cite{MPM14}). Hence the irrelevance of spin in entanglement generation starting from vacuum should be ascribed to the absence of mode mixing for the considered model of spacetime dynamics. As consequence we can argue a relevance of spin, for vacuum originated entanglement, in anisotropic spacetime where mode mixing appears.
This is an avenue to take for further studies.




\appendix

\section{Unitary representation of Bogolyubov transformations}\label{unirep}

Below we will derive the unitary operator (acting on the Fock space) that represents the transformation
\eqref{Bab} (and consequently of \eqref{Babsimple}, \eqref{ainoutspinless}).
In fact the coefficients of this latter, once arranged in matrix form, can be regarded as an element of a matrices group \cite{NNBog}.

Let us start considering fermionic mode operators  ${\sf f}_i$, satisfying the algebra
\b\label{alg}
\left\{ {\sf f}_i,{\sf f}_j^{\dagger}\right\}=\delta_{ij}\qquad
\left\{ {\sf f}_i,{\sf f}_j\right\}=0.
\e
For our purposes ${\sf f}_1:=a_{out}(\u)$,  ${\sf f}_2:=a_{out}(\d)$,  ${\sf f}_3:=b_{out}(\u)$,  and ${\sf f}_4:=b_{out}(\d)$.
Let us further define the unitary operator (fermionic `squeezing' operator)
\b\label{calU}
{\cal U}:=\exp\left[\frac{1}{2}\sum_{i_1i_2}\left(\theta_{i_1i_2}{\sf f}^{\dagger}_{i_1}{\sf f}^{\dagger}_{i_2}
+\theta^*_{i_1i_2}{\sf f}_{i_1}{\sf f}_{i_2}\right)\right],
\e
by means of complex coefficients $\theta_{i_1i_2}$ that have to be intended as entries of an antisymmetric matrix
$\theta$ with null diagonal elements.

We want to obtain the input ($in$) operators by the action of $\cal U$ on the output ($out$) ones. For example we want $a_{in}(\u)={\cal U}\, a_{out}(\u)\,{\cal U}^\dag$.
Let us then compute
\begin{equation*}
{\cal U}\,{\sf f}_j\,{\cal U}^{\dagger}={\sf f}_j+\frac{1}{1!}[L_{\cal U},{\sf f}_j]+\frac{1}{2!}[L_{\cal U},[L_{\cal U},{\sf f}_j]]+\frac{1}{3!}[L_{\cal U},[L_{\cal U}[L_{\cal U},{\sf f}_j]]]+\ldots
\end{equation*}
where
\begin{equation*}
L_{\cal U}:=\frac{1}{2}\sum_{i_1i_2}\left(\theta_{i_1i_2}{\sf f}^{\dagger}_{i_1}{\sf f}^{\dagger}_{i_2}
+\theta^*_{i_1i_2}{\sf f}_{i_1}{\sf f}_{i_2}\right)\,.
\end{equation*}
We will have
\begin{align}\label{relD}
[L_{\cal U},{\sf f}_j]&=\sum_i\theta_{ij}{\sf f}^{\dagger}_i, \notag  \\
[L_{\cal U}[L_{\cal U},{\sf f}_j]]&=\sum_{ii_1}\theta_{ij}\theta^*_{i_1i}{\sf f}_{i_1}, \notag  \\
[L_{\cal U}[L_{\cal U}[L_{\cal U},{\sf f}_j]]]&=\sum_{ii_1i_2}\theta_{i_1j}\theta^*_{ii_1}\theta_{i_2i}{\sf f}^{\dagger}_{i_2}, \notag  \\
[L_{\cal U}[L_{\cal U}[L_{\cal U}[L_{\cal U},{\sf f}_j]]]]&=\sum_{ii_1i_2i_3}\theta_{i_1j}\theta^*_{ii_1}\theta_{i_2i}\theta^*_{i_3i_2}{\sf f}_{i_3}.
\end{align}
To show how to derive such relations we provide the explicit calculation for the first of them, namely
\begin{align*}
[L_{\cal U},{\sf f}_j]&=\frac{1}{2}\sum_{i_1i_2}\left(\theta_{i_1i_2}[{\sf f}^{\dagger}_{i_1}{\sf f}^{\dagger}_{i_2},{\sf f}_j]+\theta^*_{i_1i_2}[{\sf f}_{i_1}{\sf f}_{i_2},{\sf f}_j] \right)  \\
&=\frac{1}{2}\sum_{i_1i_2}\theta_{i_1i_2} \left({\sf f}^{\dagger}_{i_1}[{\sf f}^{\dagger}_{i_2},{\sf f}_{j}]+[{\sf f}^{\dagger}_{i_1},{\sf f}_{j}]{\sf f}^{\dagger}_{i_2} \right)  \\
&=\frac{1}{2}\sum_{i_1i_2}\theta_{i_1i_2} \left({\sf f}^{\dagger}_{i_1}\{{\sf f}^{\dagger}_{i_2},{\sf f}_{j}\}-\{{\sf f}^{\dagger}_{i_1},{\sf f}_{j}\}{\sf f}^{\dagger}_{i_2} \right)    \\
&=\frac{1}{2}\sum_{i_1i_2}\theta_{i_1i_2} \left({\sf f}^{\dagger}_{i_1}\delta_{i_2j}-{\sf f}^{\dagger}_{i_2}\delta_{i_1j}\right)    \\
&=\frac{1}{2}\left(\sum_{i_1}\theta_{i_1j}{\sf f}^{\dagger}_{i_1}-\sum_{i_2}{\sf f}^{\dagger}_{i_2}\theta_{ji_2}\right)    \\
&=\sum_{i}\theta_{ij}{\sf f}^{\dagger}_{i}  \,.
\end{align*}
The relations \eqref{relD} give
\begin{align*}
{\cal U}\,{\sf f}_j\,{\cal U}^{\dagger}=&{\sf f}_j+\frac{1}{1!}\sum_{i}\theta_{ij}{\sf f}^{\dagger}_{i}+\frac{1}{2!}\sum_{ii_1}\theta_{ij}\theta^*_{i_1i}\,{\sf f}_{i_1}
+\frac{1}{3!}\sum_{ii_1i_2}\theta_{i_1j}\theta^*_{ii_1}\theta_{i_2i}\,{\sf f}^{\dagger}_{i_2}
+\frac{1}{4!}\sum_{ii_1i_2i_3}\theta_{i_1j}\theta^*_{ii_1}\theta_{i_2i}\theta^*_{i_3i_2}\,{\sf f}_{i_3} \\
&+\frac{1}{5!}\sum_{ii_1i_2i_3i_4}\theta_{i_1j}\theta^*_{ii_1}\theta_{i_2i}\theta^*_{i_3i_2}\theta_{i_4i_3}\,{\sf f}^{\dagger}_{i_4}
+\frac{1}{6!}\sum_{ii_1i_2i_3i_4i_5}\theta_{i_1j}\theta^*_{ii_1}\theta_{i_2i}\theta^*_{i_3i_2}\theta_{i_4i_3}\theta^*_{i_5i_4}\,{\sf f}_{i_5}+\ldots
\end{align*}
Swapping the indexes on $\theta$ and renaming some of them allows us to write
\begin{align}\label{Umunu}
{\cal U}\,{\sf f}_j\,{\cal U}^{\dagger}=\sum_i\Big( &{\sf f}_i\delta_{ij}-\frac{1}{1!}\theta_{ji}{\sf f}^{\dagger}_{i}-\frac{1}{2!}\sum_{i_1}\theta_{ji_1}\theta^*_{ii_1}\,{\sf f}_{i}
+\frac{1}{3!}\sum_{i_1i_2}\theta_{ji_1}\theta^*_{i_2i_1}\theta_{i_2i}\,{\sf f}^{\dagger}_{i}
+\frac{1}{4!}\sum_{i_1i_2i_3}\theta_{ji_1}\theta^*_{i_2i_1}\theta_{i_2i_3}\theta^*_{ii_3}\,{\sf f}_{i} \notag\\
&-\frac{1}{5!}\sum_{i_1i_2i_3i_4}\theta_{ji_1}\theta^*_{i_2i_1}\theta_{i_2i_3}\theta^*_{i_4i_3}\theta_{i_4i}\,{\sf f}^{\dagger}_{i}-\frac{1}{6!}\sum_{i_1i_2i_3i_4i_5}\theta_{ji_1}\theta^*_{i_2i_1}\theta_{i_2i_3}\theta^*_{i_4i_3}\theta_{i_4i_5}\theta^*_{ii_5}\,{\sf f}_{i}+\ldots \Big)   \notag   \\
&\hspace{-1.2cm}\equiv\sum_i\big(\mu_{ji}\,{\sf f}_i+\nu_{ji}\,{\sf f}^{\dagger}_i\big).
\end{align}
To determine the coefficients $\mu_{ji}$ and $\nu_{ji}$ we consider
the polar decomposition $\theta=\Theta\,|\theta|$ where $\Theta$ is a unitary matrix
and $|\theta|$ is defined through the relation $\theta^\dag\theta=|\theta|^2$.
Hence, we have
\begin{align}\label{theta2&4}
&\sum_{i_1}\theta_{ji_1}\theta_{ii_1}^*=(\theta\theta^{\dagger})_{ji}=(|\theta|^2)_{ji},  \notag \\
&\sum_{i_1i_2i_3}\theta_{ji_1}\theta_{i_2i_1}^*\theta_{i_2i_3}\theta_{ii_3}^*=\sum_{i_2}(\theta\theta^{\dagger})_{ji_2}(\theta\theta^{\dagger})_{i_2i} =\sum_{i_2}(|\theta|)_{ji_2}^2(|\theta|)_{i_2i}^2=(|\theta|^2|\theta|^2)_{ji}=(|\theta|^4)_{ji} \,,
\end{align}
and so on. Additionally, we can write
\b\label{theta1}
\theta_{ji}=\sum_{l}\delta_{jl}\theta_{li}=\sum_{l}(|\theta||\theta|^{-1})_{jl}\theta_{li}=\sum_{kl}|\theta|_{jk}|\theta|^{-1}_{kl}\theta_{li} \,.
\e
Then, using \eqref{theta2&4} and \eqref{theta1} into \eqref{Umunu}, we arrive at
\begin{align}\label{munuvsth}
\mu_{ji}&=[\cos(|\theta|)]_{ji}  \,,\notag\\
\nu_{ji}&=-\sum_{kl}[\sin(|\theta|)]_{jk}|\theta|^{-1}_{kl}\theta_{li}=-[\sin(|\theta|)|\theta|^{-1}\,\theta]_{ji} \,.
\end{align}
The coefficients $\mu_{ji}$ and $\nu_{ji}$, have to satisfy the following conditions, coming from the algebra of fermionic modes \eqref{alg},
\begin{align}\label{munucond}
\Big\{ {\sf f}_i,{\sf f}_j^{\dagger}\Big\}=\delta_{ij}\qquad &\Rightarrow \qquad\sum_{l}(\mu_{il}\mu^*_{jl}+\nu_{il}\nu^*_{jl}) = \delta_{ij}  \,,   \notag\\
\Big\{ {\sf f}_i,{\sf f}_j\Big\}=0\qquad &\Rightarrow \qquad\sum_{l}(\mu_{il}\nu_{jl}+\nu_{il}\mu_{jl}) = 0  \,.
\end{align}
Now, we write the Bogolyubov transformations \eqref{Bab} as
\begin{equation*}
\left[
\begin{array}{c}
 a_{in}(\u) \\
 a_{in}(\d)  \\
  b_{in}(\u)   \\
  b_{in}(\d)
\end{array}%
\right]=\left[%
\begin{array}{cccc}
 \mathcal A & 0 & 0 & 0 \\
  0 &   \mathcal A & 0 & 0 \\
  0 & 0 & \mathcal A & 0   \\
  0 & 0 & 0 & \mathcal A
\end{array}%
\right] \left[%
\begin{array}{c}
{\sf f}_1 \\
{\sf f}_2  \\
{\sf f}_3   \\
{\sf f}_4
\end{array}\right]
-
\left[%
\begin{array}{cccc}
 0 & 0 & -\beta^*_{\u\u} & -\beta^*_{\u\d} \\
 0 &  0  & -\beta^*_{\d\u} & -\beta^*_{\d\d} \\
\beta^*_{\u\u} & \beta^*_{\d\u} & 0 & 0   \\
\beta^*_{\u\d} & \beta^*_{\d\d} & 0 & 0
\end{array}%
\right] \left[%
\begin{array}{c}
{\sf f}_1^\dag \\
{\sf f}_2^\dag \\
{\sf f}_3^\dag   \\
{\sf f}_4^\dag
\end{array}
\right]
\end{equation*}
therefore, the matrices $\mu$ and $\nu$ have to be
\begin{equation}\label{munumatrix}
\mu=\left[%
\begin{array}{cccc}
 \mathcal A & 0 & 0 & 0 \\
  0 &   \mathcal A  & 0 & 0 \\
  0 & 0 & \mathcal A & 0   \\
  0 & 0 & 0 & \mathcal A    \\
\end{array}%
\right], \qquad\qquad
\nu=-\left[%
\begin{array}{cccc}
 0 & 0 & -\beta^*_{\u\u} & -\beta^*_{\u\d} \\
 0 &  0  &- \beta^*_{\d\u} &- \beta^*_{\d\d} \\
 \beta^*_{\u\u} &  \beta^*_{\d\u} & 0 & 0   \\
 \beta^*_{\u\d} &  \beta^*_{\d\d} & 0 & 0    \\
\end{array}%
\right] .
\end{equation}
In passing, it is worth noticing that the block matrix
$\left[\begin{array}{cc}
\mu & \nu \\
\nu^* & \mu^*
\end{array}\right]$
can be regarded as an element of a matrices group \cite{NNBog}, whose representation in the Fock space
is given by $\cal U$ of \eqref{calU}.


The conditions \eqref{munucond}, given \eqref{munumatrix},  can be translated into
\begin{align}\label{bogcoef-rel1}
\begin{cases}
|\mathcal A|^2+|\beta_{\u\u}|^2+|\beta_{\d\u}|^2 &=1   \\
|\mathcal A|^2+|\beta_{\u\d}|^2+|\beta_{\d\d}|^2 &=1  \\
|\mathcal A|^2+|\beta_{\u\u}|^2+|\beta_{\u\d}|^2 &=1   \\
|\mathcal A|^2+|\beta_{\d\d}|^2+|\beta_{\d\u}|^2 &=1
\end{cases}
\Longrightarrow
\begin{cases}
|\beta_{\u\u}|&=|\beta_{\d\d}|   \\
|\beta_{\u\d}|&=|\beta_{\d\u}|
\end{cases}\,,
\end{align}
and into
\begin{align}\label{bogcoef-rel2}
\begin{cases}
\beta_{\u\u}\beta^*_{\u\d}+\beta_{\d\u}\beta_{\d\d}^*&=0    \\
\beta_{\u\u}\beta^*_{\d\u}+\beta_{\u\d}\beta_{\d\d}^*&=0
\end{cases}\,.
\end{align}
Notice that because of \eqref{bogcoef-rel1} we have
\begin{equation*}
n^p(\u)=n^p(\d)=n^a(\u)=n^a(\d)\,.
\end{equation*}
Furthermore, from \eqref{bogcoef-rel2} we can derive the following relation
\begin{align*}\label{rel5}
\beta^{\ast}_{\downarrow\uparrow} \beta^{\ast}_{\uparrow\downarrow}-\beta^{\ast}_{\downarrow\downarrow}\beta^{\ast}_{\uparrow\uparrow}&=-|\beta_{\u\d}|^2\frac{\beta_{\d\d}^*}{\beta_{\u\u}}+|\beta_{\u\u}|^2\frac{\beta_{\u\d}^*}{\beta_{\d\u}} \notag\\
&=|\beta_{\u\d}|^2\frac{\beta_{\u\d}^*}{\beta_{\d\u}}+|\beta_{\u\u}|^2\frac{\beta_{\u\d}^*}{\beta_{\d\u}}=(|\beta_{\u\d}|^2+|\beta_{\u\u}|^2)\frac{\beta_{\u\d}^*}{\beta_{\d\u}} \notag \\
&=(|\beta_{\u\d}|^2+|\beta_{\u\u}|^2) e^{i\phi} \,.
\end{align*}


Coming back to the unitary operator $\cal U$ of \eqref{calU}, we can argue for $\theta$ the following form
\begin{equation}\label{thetamatrix1}
\theta=\left[
\begin{array}{cccc}
 0 & 0 & \vartheta_1 & \vartheta_2 \\
 0 &  0  & \vartheta_4 & \vartheta_3 \\
 -  \vartheta_1 & -  \vartheta_4 & 0 & 0   \\
 -  \vartheta_2 & -  \vartheta_3 & 0 & 0    \\
\end{array}
\right] ,
\end{equation}
and impose $\theta^\dag\theta=|\theta|^2$ to be diagonal (because $\mu$ is also diagonal).
This leads to
\begin{equation*}
\vartheta_1^*\vartheta_4+\vartheta_2^*\vartheta_3=0, \qquad \vartheta_1^*\vartheta_2+\vartheta_4^*\vartheta_3=0,
\end{equation*}
which give $|\vartheta_1|^2=|\vartheta_3|^2$ and $|\vartheta_2|^2=|\vartheta_4|^2$.
Thus
\begin{equation*}
|\theta|=diag\left[ \sqrt{| \vartheta_1|^2+| \vartheta_2|^2},    \sqrt{| \vartheta_1|^2+| \vartheta_2|^2}\,,
  \sqrt{| \vartheta_1|^2+| \vartheta_2|^2},    \sqrt{| \vartheta_1|^2+| \vartheta_2|^2}\right] \,.
\end{equation*}
As a consequence of \eqref{munuvsth} we have
\begin{equation}\label{munutheta}
\mu= \cos\left(  \sqrt{| \vartheta_1|^2+| \vartheta_2|^2}\right)
\left[
\begin{array}{cccc}
1 & 0 & 0 & 0 \\
0 & 1 & 0 & 0 \\
0 & 0 & 1 & 0   \\
0 & 0 & 0 & 1   \\
\end{array}
\right] \,,
\qquad
\nu=-\frac{\sin\left(  \sqrt{| \vartheta_1|^2+| \vartheta_2|^2}\right)}{  \sqrt{| \vartheta_1|^2+| \vartheta_2|^2}}
\left[
\begin{array}{cccc}
 0 & 0 & \vartheta_1 & \vartheta_2 \\
 0 &  0  & \vartheta_4 & \vartheta_3 \\
-  \vartheta_1 & -  \vartheta_4 & 0 & 0   \\
 -  \vartheta_2 & -  \vartheta_3 & 0 & 0    \\
\end{array}
\right] \,,
\end{equation}
with
\begin{equation}
\label{thetamatrix}
\Theta=\frac{1}{  \sqrt{| \vartheta_1|^2+| \vartheta_2|^2}}
\left[
\begin{array}{cccc}
 0 & 0 & \vartheta_1 & \vartheta_2 \\
 0 &  0  & \vartheta_4 & \vartheta_3 \\
-  \vartheta_1 & -  \vartheta_4 & 0 & 0   \\
 -  \vartheta_2 & -  \vartheta_3 & 0 & 0    \\
\end{array}
\right].
\end{equation}
Finally equating \eqref{munutheta} to \eqref{munumatrix} we will get $\cal A$ real and
\begin{align*}
\vartheta_1&=-\frac{\arccos{\cal A}}{\sin\left(\arccos{\cal A}\right)}\beta_{\u\u}^*, \\
\vartheta_2&=-\frac{\arccos{\cal A}}{\sin\left(\arccos{\cal A}\right)}\beta_{\u\d}^*, \\
\vartheta_3&=-\frac{\arccos{\cal A}}{\sin\left(\arccos{\cal A}\right)}\beta_{\d\d}^*, \\
\vartheta_4&=-\frac{\arccos{\cal A}}{\sin\left(\arccos{\cal A}\right)}\beta_{\d\u}^*.
\end{align*}


\bigskip


Here, we want to consider Bogolyubov transformations able to preserve angular momentum.
They take the following form (see Ref. \cite{MPM14})
\begin{equation*}
\left[
\begin{array}{c}
a(\u)_{in} \\
a(\d)_{in}  \\
 b(\u)_{in}   \\
b(\d)_{in}
\end{array}
\right]=\left[
\begin{array}{cccc}
 \mathcal A & 0 & 0 & 0 \\
  0 &   \mathcal A & 0 & 0 \\
  0 & 0 & \mathcal A& 0   \\
  0 & 0 & 0 & \mathcal A
\end{array}
\right]
-
\left[
\begin{array}{c}
 {\sf f}_{1} \\
 {\sf f}_{2} \\
 {\sf f}_{3} \\
 {\sf f}_{4}
\end{array}\right]
\left[
\begin{array}{cccc}
 0 & 0 & 0 & -\beta^*_{\u\d} \\
 0 &  0  & -\beta^*_{\d\u} & 0 \\
0 & \beta^*_{\d\u} & 0 & 0   \\
\beta^*_{\u\d} & 0 & 0 & 0    \\
\end{array}
\right] \left[
\begin{array}{c}
 {\sf f}_{1}^{\dagger} \\
 {\sf f}_{2}^{\dagger}  \\
 {\sf f}_{3}^{\dagger}   \\
 {\sf f}_{4}^{\dagger}
\end{array}
\right]\,,
\end{equation*}
therefore, the matrices $\mu$ and $\nu$ have to be
\begin{equation*}
\mu=\left[
\begin{array}{cccc}
 \mathcal A & 0 & 0 & 0 \\
  0 &   \mathcal A  & 0 & 0 \\
  0 & 0 & \mathcal A & 0   \\
  0 & 0 & 0 & \mathcal A    \\
\end{array}
\right], \qquad\qquad
\nu=-\left[
\begin{array}{cccc}
 0 & 0 & 0 & -\beta^*_{\u\d} \\
 0 &  0  & -\beta^*_{\d\u} & 0 \\
0 &  \beta^*_{\d\u} & 0 & 0   \\
  \beta^*_{\u\d} & 0 & 0 & 0    \\
\end{array}
\right].
\end{equation*}
This implies that the matrix \eqref{thetamatrix1} simply becomes
\begin{equation}\label{theta2}
\theta=\left[
\begin{array}{cccc}
 0 & 0 & 0 & \vartheta_2 \\
 0 &  0  & \vartheta_4 & 0 \\
0 & - \vartheta_4 & 0 & 0   \\
 - \vartheta_2 & 0 & 0 & 0    \\
\end{array}
\right] ,
\end{equation}
with $ |\vartheta_2|=|\vartheta_4| $ and $|\theta|=diag[|\vartheta_2|]$.
As consequence it is
\begin{equation*}
\Theta=\frac{1}{|\vartheta_2|}\theta,
\end{equation*}
and also
\begin{equation*}
\mu= \cos(|\vartheta_2|)\left[
\begin{array}{cccc}
1 & 0 & 0 & 0 \\
0 & 1 & 0 & 0 \\
0 & 0 & 1 & 0   \\
0 & 0 & 0 & 1   \\
\end{array}
\right],
\qquad
\nu=-\frac{\sin(|\vartheta_2|)}{|\vartheta_2|}\left[
\begin{array}{cccc}
 0 & 0 & 0 & \vartheta_2 \\
 0 &  0  & \vartheta_4 & 0 \\
0 & - \vartheta_4 & 0 & 0   \\
- \vartheta_2 & 0 & 0 & 0    \\
\end{array}
\right].
\end{equation*}

\bigskip

Finally, for the spinless transformation \eqref{ainoutspinless},
the matrix \eqref{thetamatrix} simply becomes
\begin{equation}
\label{Theta3}
\Theta=\frac{1}{|\vartheta|}\left[\begin{array}{cc}
0 & \vartheta \\
-\vartheta & 0
\end{array}\right],
\end{equation}
with $\vartheta=-\frac{\arccos{\cal A}}{\sin(\arccos{\cal A})}\beta^*$, and
the unitary operator reads
\begin{equation}
{\cal U}= \exp\left[  \vartheta \,a_{out}^\dag b_{out}^\dag + \vartheta^* \,a_{out} b_{out} \right].
\label{Uspinless}
\end{equation}


\section{Decoupling the unitary operator $\cal U$}\label{Uact}

In this Appendix we will show how the unitary operator defined in \eqref{calU} acts on a vacuum state.
First of all it is useful to recall the following result for the exponentiation of operators belonging to the
${\mathfrak{su}}(2)$ algebra \cite{Wilcox}.

Given operators $K_\pm$ and $K_3$ satisfying the following commutation relations
\begin{equation*}
[K_+,K_-]=-2K_3,
\quad\quad
[K_3,K_{\pm}]=\pm K_{\pm},
\end{equation*}
it is
\b
\label{dectheo1}
e^{(\gamma_+K_++\gamma_-K_-+\gamma_3K_3)}=e^{ \Gamma_+K_+}e^{(\ln\Gamma_3)K_3}e^{\Gamma_-K_-} \,,
\e
with
\begin{equation*}
\Gamma_{\pm}\equiv\frac{2\gamma_{\pm}\sinh\xi}{2\xi\cosh\xi-\gamma_3\sinh\xi},
\quad
\Gamma_3\equiv\Big(\cosh\xi -\frac{\gamma_3}{2\xi\sinh\xi}\Big)^{-2},
\quad
\xi^2\equiv\frac{1}{4}\gamma_3^2-\gamma_+\gamma_-.
\end{equation*}

For our purposes we take
\begin{align*}\label{Kpm}
K_+&=\frac{1}{2r}\sum_{i_1i_2}\theta_{i_1i_2}{\sf f}^{\dagger}_{i_1}{\sf f}^{\dagger}_{i_2}\,,  \nonumber\\
K_-&=\frac{1}{2r}\sum_{i_1i_2}\theta^*_{i_1i_2}{\sf f}_{i_1}{\sf f}_{i_2} \,,
\end{align*}
then
$\gamma_{\pm}=r$ and $\gamma_3=0$, with $r$ the eigenvalue of the matrix $|\theta|$. It also results
$\beta=\pm i r$,
$\Gamma_3=\frac{1}{\cos^2r}$,
$ \Gamma_{\pm}=\tan r $.

Furthermore we take
\begin{equation*}
K_3=\frac{1}{4}\sum_{i_2}\Big(2{\sf f}^{\dagger}_{i_2}{\sf f}_{i_2}-4 I\Big) \,,
\end{equation*}
where $I$ is the identity operator.

It can be easily shown that with our choice of the operators $K_\pm$ and $K_3$
we are allowed to write the exponential \eqref{calU} as
\begin{align}\label{Udecomp}
&\exp\left\{\frac{1}{2}\sum_{i_1i_2}\left[\theta_{i_1i_2}{\sf f}^{\dagger}_{i_1}{\sf f}^{\dagger}_{i_2}
+\theta^*_{i_1i_2}{\sf f}_{i_2}{\sf f}_{i_1}\right] \right\} \nonumber\\
&=\exp\left[\frac{\tan r}{2r}\sum_{i_1i_2}\theta_{i_1i_2}{\sf f}^{\dagger}_{i_1}{\sf f}^{\dagger}_{i_2}\right]
\left\{(\cos^2 r) \, \exp\left[\frac{-\ln(\cos^2 r )}{4}\sum_{i_1}2{\sf f}^{\dagger}_{i_1}{\sf f}_{i_1}\right]\right\}
\exp\left[\frac{\tan r}{2r}\sum_{i_1i_2}\theta^*_{i_1i_2}{\sf f}_{i_1}{\sf f}_{i_2}\right] \,.
\end{align}

Hence when applying the unitary operator ${\cal U}$ to the vacuum we are left with
\b
\label{uaction}
{\cal U}|0_p;0_{-p}\rangle_{out}=\cos^2 r \sum_n\frac{\tan^n r}{(r)^n n!}\left(\frac{1}{2}\sum_{i_1i_2}\theta_{i_1i_2}
{\sf f}^{\dagger}_{i_1}{\sf f}^{\dagger}_{i_2}\right)^n |0_p;0_{-p}\rangle_{out}  \,.
\e

Considering Eq.\eqref{thetamatrix1} and ${\sf f}^{\dagger}_i=\{a^{\dagger}_{out}(\u),\,a_{out}^{\dagger}(\d),\,b_{out}^{\dagger}(\u),\,b_{out}^{\dagger}(\d)\}$ give, in the most general case we considered
\begin{align}\label{cre-fact}
&\left(\frac{1}{2}\sum_{i_1i_2}\theta_{i_1i_2}{\sf f}^{\dagger}_{i_1}{\sf f}^{\dagger}_{i_2}\right)
=\vartheta_1a_{out}^{\dagger}(\u)b_{out}^{\dagger}(\u)+\vartheta_2a_{out}^{\dagger}(\u)b_{out}^{\dagger}(\d)+\vartheta_4a_{out}^{\dagger}(\d)b_{out}^{\dagger}(\u)+\vartheta_3a_{out}^{\dagger}(\d)b_{out}^{\dagger}(\d), \nonumber \\
&\left(\frac{1}{2}\sum_{i_1i_2}\theta_{i_1i_2}{\sf f}^{\dagger}_{i_1}{\sf f}^{\dagger}_{i_2}\right)^2
=2(\vartheta_2\vartheta_4-\vartheta_1\vartheta_3)
a_{out}^{\dagger}(\u)a_{out}^{\dagger}(\d)b_{out}^{\dagger}(\u)b_{out}^{\dagger}(\d), \\
&\left(\frac{1}{2}\sum_{i_1i_2}\theta_{i_1i_2}{\sf f}^{\dagger}_{i_1}{\sf f}^{\dagger}_{i_2}\right)^n=0,\quad n>2  \nonumber   \,.
\end{align}

The expression \eqref{uaction} can now be readily computed as
\begin{align}\label{Uaction}
{\cal U}|0_p;0_{-p}\rangle_{out}&=\cos^2r\,\Big\{1+\frac{\tan r}{r}\big[\vartheta_1a_{out}^{\dagger}(\u)b_{out}^{\dagger}(\u)+\vartheta_2a_{out}^{\dagger}(\u)b_{out}^{\dagger}(\d)+\vartheta_4a_{out}^{\dagger}(\d)b_{out}^{\dagger}(\u)+\vartheta_3a_{out}^{\dagger}(\d)b_{out}^{\dagger}(\d)\big]  \nonumber\\
&\hspace{0.5cm}
+\frac{\tan^2 r}{r^2}(\vartheta_2\vartheta_4-\vartheta_1\vartheta_3)a_{out}^{\dagger}(\u)a_{out}^{\dagger}(\d)b_{out}^{\dagger}(\u)b_{out}^{\dagger}(\d) \Big\}|0\rangle_{out} \nonumber\\
&=\mathcal{A}^2\Big\{|0_p;0_{-p}\rangle_{out}+\frac{1}{\mathcal A}\big[\beta^*_{\u\u}|\u_p;\u_{-p}\rangle_{out}+\beta^*_{\u\d}|\u_p;\d_{-p}\rangle_{out}+\beta^*_{\d\u}|\d_p;\u_{-p}\rangle_{out}+\beta^*_{\d\d}|\d_p;\d_{-p}\rangle_{out} \big]  \\
&\hspace{0.5cm}
+\frac{1}{\mathcal{A}^2}(\beta^*_{\u\d}\beta^*_{\d\u}-\beta^*_{\u\u}\beta^*_{\d\d})|\u\d_p;\u\d_{-p}\rangle_{out} \Big\}  \nonumber\,,
\end{align}
where we have used the fact that $r=\sqrt{|\vartheta_1|^2+|\vartheta_2|^2}$ and hence
$\cos r=\mathcal A$.

\bigskip

In case of populated in-states we have also to account for the action of the first and second term in Eq.\eqref{Udecomp}. Actually the first term can be computed with relation analogous to \eqref{cre-fact} having creation operators replaced by annihilation ones, while the second term (inside curly brackets) can be understood by noticing that
\begin{align}\label{Nopseries}
(\cos^2 r)\sum_n\frac{-\ln^n(\cos^2 r)}{2^n n!}\left(\sum_{i_1}{\sf f}^{\dagger}_{i_1}{\sf f}_{i_1}\right)^n |\psi\re_{out}
=\frac{1}{\cos^{N-2} r}|\psi\re_{out},  
\end{align}
where $N$ is the number of particles-antiparticles in the state $|\psi\re_{out}$.
Thus, using \eqref{Udecomp} and the position $|\psi\re_{in}={\cal U}|\psi\re_{out}$, we get the following excited states in the out region:
\begin{align}\label{excitedC}
|\u\d_p ; \u\d_{-p}\rangle_{in}&=(\beta_{\u\d}\beta_{\d\u}-\beta_{\u\u}\beta_{\d\d})|0_p;0_{-p}\rangle_{out}+\mathcal A\beta_{\d\u}|\u_p;\d_{-p}\rangle_{out}+\mathcal A\beta_{\u\d}|\d_p;\u_{-p}\rangle_{out}-\mathcal A\beta_{\d\d}
|\u_p;\u_{-p}\rangle_{out}  \notag\\
&-\mathcal A\beta_{\u\u}|\d_p;\d_{-p}\rangle_{out}+\mathcal{A}^2|\u\d_p;\u\d_{-p}\rangle_{out}, \notag\\
|\u\d_p ; \u_{-p}\rangle_{in}&=\beta_{\d\u}|\u_p;0_{-p}\rangle_{out}-\beta_{\u\u}|\d_p;0_{-p}\rangle_{out}+\mathcal A|\u\d_p;\u_{-p}\rangle_{out}, \notag\\
|\u_p ; 0_{-p}\rangle_{in}&=\mathcal A|\u_p;0_{-p}\rangle_{out}-\beta^*_{\d\u}|\u\d_p;\u_{-p}\rangle_{out}-\beta^*_{\d\d}|\u\d_p;\d_{-p}\rangle_{out}, \notag \\
|0_p ; \u_{-p}\rangle_{in}&=\mathcal A|0_p;\u_{-p}\rangle_{out}+\beta^*_{\d\d}|\d_p;\u\d_{-p}\rangle_{out}+\beta^*_{\u\d}|\u_p;\u\d_{-p}\rangle_{out}, \\
|\u\d_p ; 0_{-p}\rangle_{in}&=|\u\d_p ; 0_{-p}\rangle_{out},\notag\\
|\u_p ; \u_{-p}\rangle_{in}&=\mathcal A\beta_{\u\u}|0_p;0_{-p}\rangle_{out}-\beta_{\u\u}\beta^*_{\u\d}|\u_p;\d_{-p}\rangle_{out}+\beta_{\u\d}\beta^*_{\d\d}|\d_p;\u_{-p}\rangle_{out}+(\mathcal A^2+|\beta_{\u\d}|^2)|\u_p;\u_{-p}\rangle_{out} \notag \\
&-\beta_{\u\u}\beta^*_{\d\d}|\d_p;\d_{-p}\rangle_{out}+\mathcal A\beta^*_{\d\d}|\u\d_p;\u\d_{-p}\rangle_{out}, 
\notag\\
|\d_p ; \u_{-p}\rangle_{in}&=\mathcal A\beta_{\d\u}|0_p;0_{-p}\rangle_{out}+(\mathcal A^2+|\beta_{\d\d}|^2)|\d_p;\u_{-p}\rangle_{out}-\beta^*_{\u\d}\beta_{\d\u}|\u_p;\d_{-p}\rangle_{out}+\beta_{\d\d}\beta^*_{\u\d}|\u_p;\u_{-p}\rangle_{out} \notag\\
&-\beta_{\d\u}\beta^*_{\d\d}|\d_p;\d_{-p}\rangle_{out}-\mathcal A\beta^*_{\u\d}|\u\d_p;\u\d_{-p}\rangle_{out}.\notag
\end{align}

\bigskip

In case of angular momentum conservation  Eq.\eqref{theta2} and ${\sf f}^{\dagger}_i=\{a^{\dagger}_{out}(\u),\,a_{out}^{\dagger}(\d),\,b_{out}^{\dagger}(\u),\,b_{out}^{\dagger}(\d)\}$ give
\begin{align}\label{cre-mom}
&\left(\frac{1}{2}\sum_{i_1i_2}\theta_{i_1i_2}{\sf f}^{\dagger}_{i_1}{\sf f}^{\dagger}_{i_2}\right)=\vartheta_2a_{out}^{\dagger}(\u)b_{out}^{\dagger}(\d)+\vartheta_4a_{out}^{\dagger}(\d)b_{out}^{\dagger}(\u),  \notag\\
&\left(\frac{1}{2}\sum_{i_1i_2}\theta_{i_1i_2}{\sf f}^{\dagger}_{i_1}{\sf f}^{\dagger}_{i_2}\right)^2=2\vartheta_2\vartheta_4a_{out}^{\dagger}(\u)a_{out}^{\dagger}(\d)b_{out}^{\dagger}(\u)b_{out}^{\dagger}(\d),   \\
&\left(\frac{1}{2}\sum_{i_1i_2}\theta_{i_1i_2}{\sf f}^{\dagger}_{i_1}{\sf f}^{\dagger}_{i_2}\right)^n=0, \quad n>2.\notag
\end{align}
Then  \eqref{uaction} yields
\begin{align*}
{\cal U}|0_p;0_{-p}\rangle_{out}&=\cos^2r\,\Big\{1+\frac{\tan r}{r}\big[\vartheta_2a_{out}^{\dagger}(\u)b_{out}^{\dagger}(\d)+\vartheta_4a_{out}^{\dagger}(\d)b_{out}^{\dagger}(\u)\big]\nonumber\\
&+\frac{\tan^2 r}{r^2}\vartheta_2\vartheta_4a_{out}^{\dagger}(\u)a_{out}^{\dagger}(\d)b_{out}^{\dagger}(\u)b_{out}^{\dagger}(\d) \Big\}|0\rangle_{out} \\
&=\mathcal{A}^2\Big\{|0_p;0_{-p}\rangle_{out}+\frac{1}{\mathcal A}\big[\beta^*_{\u\d}|\u_p;\d_{-p}\rangle_{out}+\beta^*_{\d\u}|\d_p;\u_{-p}\rangle_{out}\big]+\frac{1}{\mathcal{A}^2}\beta^*_{\u\d}\beta^*_{\d\u}|\u\d_p;\u\d_{-p}\rangle_{out} \Big\},
\end{align*}
where we have used the fact that $r=|\vartheta_2|$ and hence
$\cos r=\mathcal A$.

For populated in-states we have to account for the relations \eqref{cre-mom}, their adjoints, 
and for the operator \eqref{Nopseries} into \eqref{Udecomp}.
Thus, using the position $|\psi\re_{in}={\cal U}|\psi\re_{out}$, we get the following excited states in the out region:
\begin{align}\label{excitedCM}
|\u\d_p ; \u\d_{-p}\rangle_{in}&=\beta_{\u\d}\beta_{\d\u}|0_p;0_{-p}\rangle_{out}+\mathcal A\beta_{\d\u}|\u_p;\d_{-p}\rangle_{out}+\mathcal A\beta_{\u\d}|\d_p;\u_{-p}\rangle_{out} +\mathcal{A}^2|\u\d_p;\u\d_{-p}\rangle_{out},\notag \\
|\u_p ; \u_{-p}\rangle_{in}&=|\u_p;\u_{-p}\rangle_{out}, \notag \\
|\d_p ; \u_{-p}\rangle_{in}&=\mathcal A\beta_{\d\u}|0_p;0_{-p}\rangle_{out}+\mathcal A^2|\d_p;\u_{-p}\rangle_{out}-\beta^*_{\u\d}\beta_{\d\u}|\u_p;\d_{-p}\rangle_{out}-\mathcal A\beta^*_{\u\d}|\u\d_p;\u\d_{-p}\rangle_{out}, \notag \\
|\u\d_p ; \u_{-p}\rangle_{in}&=\beta_{\d\u}|\u_p;0_{-p}\rangle_{out}+\mathcal A|\u\d_p;\u_{-p}\rangle_{out},\notag \\
|\u_p ; 0_{-p}\rangle_{in}&=\mathcal A|\u_p;0_{-p}\rangle_{out}-\beta^*_{\d\u}|\u\d_p;\u_{-p}\rangle_{out}, \\
|0_p ; \u_{-p}\rangle_{in}&=\mathcal A|0_p;\u_{-p}\rangle_{out}+\beta^*_{\u\d}|\u_p;\u\d_{-p}\rangle_{out}, \notag\\
|0_p ; \d_{-p}\rangle_{in}&=\mathcal A|0_p;\d_{-p}\rangle_{out}-\beta^*_{\d\u}|\d_p;\u\d_{-p}\rangle_{out} ,\notag\\
|\u\d_p ; 0_{-p}\rangle_{in}&=|\u\d_p;0_{-p}\rangle_{out}. \notag
\end{align}

\bigskip

Finally, in the spinless case by referring to \eqref{Uspinless}
we take
\begin{eqnarray*}
K_+&=& \frac{1}{r}\,\vartheta \,a^{\dagger}_{out}b^{\dagger}_{out} \\
K_-&=&   \frac{1}{r}\,\vartheta^* \,a_{out}b_{out} \\
K_3&=&\frac{1}{2}(a^{\dagger}_{out}a_{out}+b^{\dagger}_{out}b_{out}-I)       \,,
\end{eqnarray*}
where $r=|\vartheta|$.
Then applying \eqref{dectheo1} we get
\ba
{\cal U}|0_p;0_{-p}\re_{out}={\cal A}\left(|0_p;0_{-p}\re_{out}-\frac{\beta^*}{\cal A}|1_p;1_{-p}\re_{out}\right).
\label{invacderivation}
\ea
and, using the position $|\psi\re_{in}={\cal U}|\psi\re_{out}$, also
\begin{align}\label{excitedspinless}
&|1_p ; 1_{-p}\rangle_{in}=\beta|0\rangle_{out}+\mathcal A|1_p1_{-p}\rangle_{out},\notag \\
&|0_p ; 1_{-p}\rangle_{in}=|0_p;1_{-p}\rangle_{out}, \\
&|1_p ; 0_{-p}\rangle_{in}=|1_p;0_{-p}\rangle_{out}.\notag
\end{align}


\section{Subsystem entropies for excited states}\label{Excited}

In the case of charge conservation the entropy of the reduced density 
operators can then be straightforwardly computed from \eqref{excitedC}, and results as:
\begin{align*}
&S\left({\rm Tr}_{-p}\left[{\cal U}\,|\u\d_p ; \u\d_{-p}\rangle_{out}\le\u\d_p ; \u\d_{-p}|{\cal U}^\dag\right]\right)= -2\left(\frac{4-n}{4}\right)\log_2 \left(\frac{4-n}{4}\right)-2\left(\frac{n}{4}\right)\log_2\left(\frac{n}{4}\right), \\
&S\left({\rm Tr}_{-p}\left[\mathcal U\,|\u\d_p ; 0_{-p}\rangle_{out}\le\u\d_p ; 0_{-p}|\,\mathcal U^{\dagger}\right]\right)=0, \\
&S\left({\rm Tr}_{-p}\left[\mathcal U\,|\u_p ; 0_{-p}\rangle_{out}\le\u_p ; 0_{-p}|\,\mathcal U^{\dagger}\right]\right)= -\left(\frac{4-n}{4}\right)\log_2 \left(\frac{4-n}{4}\right)-\left(\frac{n}{4}\right)\log_2\left(\frac{n}{4}\right), \\
&S\left({\rm Tr}_{-p}\left[\mathcal U\,|\u\d_p ; \u_{-p}\rangle_{out}\le\u\d_p ; \u_{-p}|\,\mathcal U^{\dagger}\right]\right)= -\left(\frac{4-n}{4}\right)\log_2 \left(\frac{4-n}{4}\right)-\left(\frac{n}{4}\right)\log_2\left(\frac{n}{4}\right), \\
&S\left({\rm Tr}_{-p}\left[\mathcal U\,|0_p ; \u_{-p}\rangle_{out}\le0_p ; \u_{-p}\,|\mathcal U^{\dagger}\right]\right)= -\left(\frac{4-n}{4}\right)\log_2 \left(\frac{4-n}{4}\right)-\left(\frac{n}{4}\right)\log_2\left(\frac{n}{4}\right),
\end{align*}
and
\begin{align*}
& S\left({\rm Tr}_{-p}\left[\mathcal U\,|\u_p ; \u_{-p}\rangle_{out}\le\u_p ; \u_{-p}|\,\mathcal U^{\dagger}\right]\right)=- \log_2\left[(1-\lambda)\frac{n(4-n)}{16}\right] \\
&\qquad\qquad\qquad\qquad-\sqrt{1-(1-\lambda)\frac{n(4-n)}{4}}\log_2\left[1+\sqrt{1-(1-\lambda)\frac{n(4-n)}{4}}\right]  \\
&\qquad\qquad\qquad\qquad +\sqrt{1-(1-\lambda)\frac{n(4-n)}{4}}\log_2\left[1-\sqrt{1-(1-\lambda)\frac{n(4-n)}{4}}\right], \\
&S\left({\rm Tr}_{-p}\left[\mathcal U\,|\d_p ; \u_{-p}\rangle_{out}\le\d_p ; \u_{-p}|\,\mathcal U^{\dagger}\right]\right)=-\log_2\left[\lambda\frac{n(4-n)}{16}\right] \\
&\qquad\qquad\qquad\qquad-\sqrt{1-\lambda\frac{n(4-n)}{4}}\log_2\left[1+\sqrt{1-\lambda\frac{n(4-n)}{4}}\right] \\
&\qquad\qquad\qquad\qquad+\sqrt{1-\lambda\frac{n(4-n)}{4}}\log_2\left[1-\sqrt{1-\lambda\frac{n(4-n)}{4}}\right], 
\end{align*}
where $\lambda$ is a parameter entering into play because 
the modulus of beta coefficients can be directly related to the particles density 
only via  
$$
|\beta_{\u\u}|^2=(1-\lambda)\frac{n}{4}, \qquad\qquad  |\beta_{\u\d}|^2=\lambda\frac{n}{4}, \qquad\qquad 
0\leq \lambda \leq 1.
$$ 
In case of charge and angular momentum conservation, Eq.\eqref{excitedCM} provide us with
\begin{align*}
&S\left({\rm Tr}_{-p}\left[\mathcal U\,|\u\d_p ; \u\d_{-p}\rangle_{out}\le\u\d_p ; \u\d_{-p}|\,\mathcal U^{\dagger}\right]\right)= -2\left(\frac{4-n}{4}\right)\log_2 \left(\frac{4-n}{4}\right)-2\left(\frac{n}{4}\right)\log_2\left(\frac{n}{4}\right), \\
&S\left({\rm Tr}_{-p}\left[\mathcal U\,|\u\d_p ; 0_{-p}\rangle_{out}\le\u\d_p ; 0_{-p}|\,\mathcal U^{\dagger}\right]\right)=0, \\
&S\left({\rm Tr}_{-p}\left[\mathcal U\,|\u_p ; 0_{-p}\rangle_{out}\le\u_p ; 0_{-p}|\,\mathcal U^{\dagger}\right]\right)= -\left(\frac{4-n}{4}\right)\log_2 \left(\frac{4-n}{4}\right)-\left(\frac{n}{4}\right)\log_2\left(\frac{n}{4}\right), \\
&S\left({\rm Tr}_{-p}\left[\mathcal U\,|\u\d_p ; \u_{-p}\rangle_{out}\le\u\d_p ; \u_{-p}|\,\mathcal U^{\dagger}\right]\right)= -\left(\frac{4-n}{4}\right)\log_2\left(\frac{4-n}{4}\right)-\left(\frac{n}{4}\right)\log_2\left(\frac{n}{4}\right), \\
&S\left({\rm Tr}_{-p}\left[\mathcal U\,|0_p ; \u_{-p}\rangle_{out}\le0_p ; \u_{-p}|\,\mathcal U^{\dagger}\right]\right)= -\left(\frac{4-n}{4}\right)\log_2\left(\frac{4-n}{4}\right)-\left(\frac{n}{4}\right)\log_2\left(\frac{n}{4}\right), \\
&S\left({\rm Tr}_{-p}\left[\mathcal U\,|\u_p ; \u_{-p}\rangle_{out}\le\u_p ; \u_{-p}|\,\mathcal U^{\dagger}\right]\right)=0,\\   
&S\left({\rm Tr}_{-p}\left[\mathcal U\,|\d_p ; \u_{-p}\rangle_{out}\le\d_p ; \u_{-p}|\,\mathcal U^{\dagger}\right]\right)= -2\left(\frac{4-n}{4}\right)\log_2\left(\frac{4-n}{4}\right)-2\left(\frac{n}{4}\right)\log_2\left(\frac{n}{4}\right).
\end{align*}
The main difference with respect to the previous case is that here entropy only depends on the particle density $n$.
Furthermore, there is no entanglement creation from one particle and one anti-particle with both spin up (or down). This was expected because of the conservation of angular momentum.

Finally, concerning the spineless case, as consequence of \eqref{excitedspinless}, we have 
\begin{align*}
&S\left({\rm Tr}_{-p}\left[\mathcal U\,|1_p ; 1_{-p}\rangle_{out}\le 1_p ; 1_{-p}|\,\mathcal U^{\dagger}\right]\right)= -\left(\frac{2-n}{2}\right)\log_2\left(\frac{2-n}{2}\right)-\left(\frac{n}{2}\right)\log_2\left(\frac{n}{2}\right), \\
&S\left({\rm Tr}_{-p}\left[\mathcal U\,|1_p ; 0_{-p}\rangle_{out}\le1_p ; 0_{-p}|\,\mathcal U^{\dagger}\right]\right)= 0, \\
&S\left({\rm Tr}_{-p}\left[\mathcal U\,|0_p ; 1_{-p}\rangle_{out}\le0_p ; 1_{-p}|\,\mathcal U^{\dagger}\right]\right)= 0.
\end{align*}

\newpage


\begin{thebibliography}{99}

\bibitem{PT04}
A. Peres and D. R. Terno, Rev. Mod. Phys. \textbf{76}, 93 (2004).

\bibitem{Ivet-rev}
P. M. Alsing and I. Fuentes, Class. Quantum Grav. \textbf{29}, (2012)

\bibitem{exp}
D. Rideout et al., Class. Quantum Grav. \textbf{29}, 224011 (2012)

\bibitem{Horo}
R. Horodecki, P. Horodecki, M. Horodecki, and K. Horodecki, Rev. Mod. Phys. \textbf{81}, 865 (2009).

\bibitem{JSS07}
T. F. Jordan, A. Shaji and E. C. G. Sudarshan, Phys. Rev. A \textbf{75}. 022101 (2007).

\bibitem{CCM12}
C. Cafaro, S. Capozziello, S. Mancini, Int. J. Theor. Phys. \textbf{51}, 2313 (2012).

\bibitem{FS05}
I. Fuentes-Schuller and R. B. Mann, Phys. Rev. Lett. \textbf{95}, 120404 (2005).

\bibitem{AL06}
P. M. Alsing, I. Fuentes-Schuller, R. B. Mann, and T. E.
Tessier, Phys. Rev. A \textbf{74}, 032326 (2006).

\bibitem{M09}
R. B. Mann and V. M. Villalba, Phys. Rev. A \textbf{80}, 022305  (2009).

\bibitem{L09}
J. Leon and  E. M. Martinez  Phys. Rev. A \textbf{80}, 012314 (2009).

\bibitem{bdv}
N. D. Birrell and P. C. W. Davies, \textit{Quantum Fields in Curved Space}
(Cambridge University Press, Cambridge, 1982).

\bibitem{TU04}
H. Terashima and M. Ueda, Phys. Rev. A \textbf{69}, 032113 (2004).

\bibitem{Shi04}
Y. Shi, Phys. Rev. D \textbf{70}, 105001 (2004).

\bibitem{ball}
J. L. Ball, I. F. Schuller, F .P. Schuller, Phys. Lett A {\textbf{359}}, 550 (2006).

\bibitem{st}
G. Ver Steeg and  N. C. Menicucci, Phys. Rev. D {\textbf{79}}, 044027 (2009).

\bibitem{iv}
I. Fuentes, R. B. Mann, E. Martin-Martinez, S. Moradi  Phys. Rev. D {\textbf{82}}, 045030 (2010).

\bibitem{MMM12}
E. Martin-Martinez and N. C. Menicucci, Class. Quantum Grav. \textbf{29}, 224003 (2012).

\bibitem{MPM14}
S. Moradi, R. Pierini and S. Mancini, Phys. Rev. D \textbf{89}, 024022 (2014).

\bibitem{D78}
A. Duncan, Phys. Rev. D \textbf{17}, 964 (1978).

\bibitem{NNBog}
N. N. Bogolyubov, Zh. Eksp. Teo. Fiz. \textbf{34}, 58 (1958) [Sov. Phys. JETP  \textbf{34}, 41 (1958)].

\bibitem{Wilcox}
R. M. Wilcox, J. Math. Phys. \textbf{8}, 962 (1967).

\end{thebibliography}
\end{document}